\documentclass[twocolumn,10pt]{asme2e}

\usepackage{graphicx}
\usepackage{environ}

\title{RELIABILITY STUDIES OF ELECTRONIC COMPONENTS FOR THE OPERATION AT CRYOGENIC TEMPERATURE}

\author{$^{*}$N.Poonthottathil, F.Krennrich, A.Weinstein, J.Eisch
    \affiliation{
	Department of Physics and Astronomy\\
	Iowa State University\\
    Ames, Iowa, 50010\\
    Email: pnav@iastate.edu
    }	
}

\author{L.J. Bond, D. Barnard, Z. Zhang, L. Koester\thanks{Address all correspondence to this author (pnav@iastate.edu).} 
    \affiliation{Center for Non-Destructive Evaluation \\
	Iowa State University\\
	Ames, Iowa, 50010\\
	Email: dbarnard@iastate.edu
    }
}

\begin{document}

\maketitle    

\begin{abstract}
\vskip.10 in
{\it Cold electronics is a key technology in many areas of science and technology including space exploration programs and particle physics. A major experiment with a very large number of analog and digital electronics signal processing channels to be operated at cryogenic temperatures is the next-generation neutrino experiment, the Deep Underground Neutrino Experiment (DUNE). The DUNE detector uses liquid Argon at 87K as a target material for neutrinos, and as a medium to track charged particles resulting from interactions in the detector volume. The DUNE electronics [1] consists of custom-designed ASIC (Application Specific Integrated Circuits) chips based on low power 180 nm-CMOS technology. The main risk for this technology is that the electronics components will be immersed in liquid argon for many years (20-30 years) without access. Reliability issues of ASICs may arise from thermal stress, packaging, and manufacturing-related defects: if undetected those could lead to long-term reliability and performance problems. The scope of this paper is to explore non-destructive evaluation techniques for their potential use in a comprehensive quality control process during prototyping, testing and commissioning of the DUNE cold electronics system. Specifically, we have used the Scanning Acoustic Microscopy and X-ray tomography to study permanent structural changes in the ASIC chips associated with thermal cycling between the room and cryogenic temperatures.}

\end{abstract}

\begin{nomenclature}
\entry{$\phi$}{Phi.}
\end{nomenclature}
\section*{INTRODUCTION}
\vskip.10 in
Next-generation neutrino physics experiments will provide precision measurements to identify neutrino induced charged particle tracks. The Liquid-argon technology-based Time Projection Chambers (LAr-TPC) for DUNE offers a unique opportunity to address two major goals in the neutrino physics, a measurement of CP-violation and to determine the neutrino mass hierarchy [3]. A major challenge is that the detector readout has to operate for many years at cryogenic temperature with stable performance. Low noise performance in the cold and proximity of the signal pick up close to the wires favors the placement of the electronics in the liquid. A LAr-TPC contains a modular electronics system with 3,000 Front-End Motherboards (FEMBs). Each FEMB provides for the signal processing of 128-channels (wires) and may host up to 18 ASICs bringing the total number of ASICs to ~ 50,000 per LAr -TPC.  Functionally, these chips include an amplifier, signal shaping, digitizer, and data assembly functions.  Data transmission is achieved via a serial copper cold cable at transmission speeds of 1 Gb/s.  \par

CMOS technology is well suited for operation at cryogenic temperature, due to its low power usage and minimal heating of the cryogenic liquid, thereby reducing the formation of bubbles in the detector. CMOS failure mechanisms, such as hot-carrier effect, and dielectric breakdown at cryogenic temperature can be minimized by tuning the operating voltages and were discussed by [2]. Detailed lifetime studies can be found in [4][5]. 

However, the reliability of the electronics components remains a key concern. This 12-bit ADC operating at 2 Mega bytes per second at cryogenic temperature is extremely challenging given the power and area constraints. Any defects associated with the manufacturing process (bubbles in the thermal paste and plastic enclosure) and thermal stresses imposed during the cryo-cycling of the chips during testing and deployment of the detector could impact the long-term performance of the system.\par 

In this paper, we demonstrate the use of Scanning Acoustic Microscopy to probe and identify defects in the chip packaging or circuitry resulting from thermal cycling. A dedicated correlation analysis technique is presented. 
Scanning Acoustic Microscopy (SAM) is a powerful non-destructive evaluation technique for the inspection of microelectronics integrated circuit packages [6]. The main advantage of SAM is that it is highly sensitive to defects including voids, cracks, and delamination inside the packaged chip. The technique also helps to identify weak spots inside the package. The integrated circuits are made from composite materials and are susceptible to thermally induced stress. A such induced discontinuity inside the chip might affect the reliability of an integrated circuit and could cause long-term performance problems. In addition to the SAM technique, we have also used X-ray microscopy to inspect chip internal wire-bonds.  Broken wire-bonds and solder defects can be detected using X-ray microscopy. Not all defects are sensitive to both X-ray and SAM, it depends on the nature of the defects [7].
\begin{figure}[h]
\includegraphics[width=8cm]{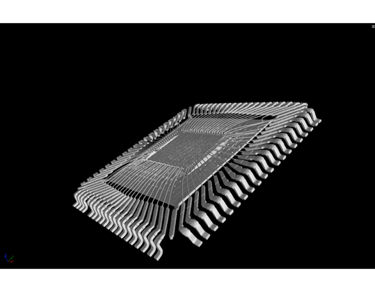}
\caption{ X-ray micrograph image of a ProtoDune ASIC, the wire-bond and the internal structure are clearly visible, the spatial resolution of the instrument is 10 $\mu m$.}
\end{figure}

\section*{EXPERIMENTAL METHODS AND MATERIALS}
\vskip.10 in
SAM uses ultrasound to interrogate a material by focusing periodic sound waves onto a small region of a sample.  The technique uses a transducer to generate, transmit and detect waves reflected by the various interfaces inside the sample. Timing and amplitude of the reflected wave provide information about changes in acoustic impedance, i.e., different materials along the path of the sound wave. The amplitude of the reflected signal is a measure of the strength of the echoes and the position on the time axis represents the time required to travel through the sample (it is a measure of the depth). Figure 3 represents an example of a reflected wave signal.

\begin{figure}[h]
\includegraphics[width=9.5cm]{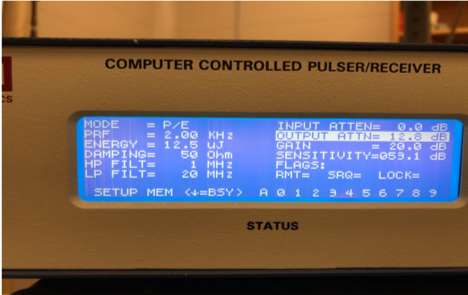}
\includegraphics[width=9.5cm]{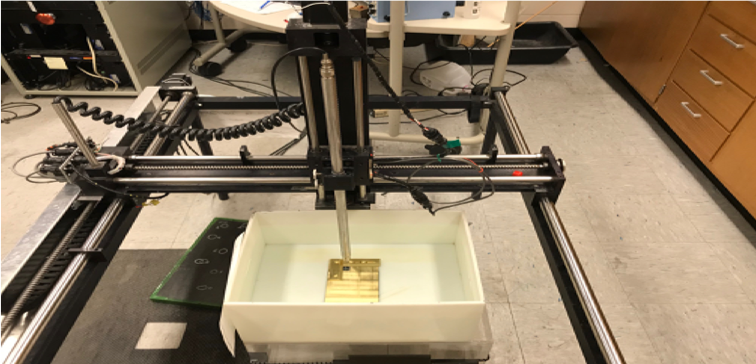}
\caption{The image shows a pulser/receiver (top), which is connected to the transducers that generates acoustic wave. A fixture  is used to place the sample that avoids the positional variations from scan to scan (bottom).}
\end{figure}

\begin{figure}[h]
\includegraphics[width=10cm]{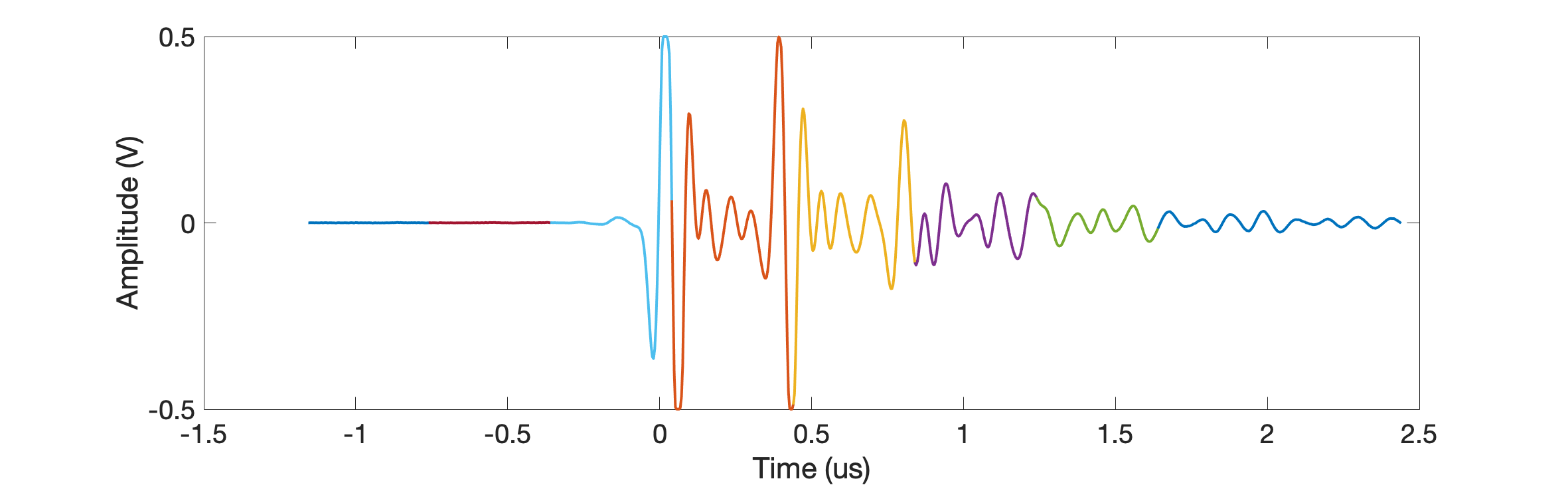}
\caption{ An example of reflected wave signal. Each color represents different depth in the sample.}
\end{figure}

For collecting the data presented in this paper, the scanning microscopy apparatus is set to its simplest configuration, called the ‘pulse-echo mode’. With this particular configuration, a single transducer generates the pulse and receives the reflected signal at a given position of the sample. The transducer sends pulses into the material using water as a coupling medium and collects the reflected signal. The signal will be a function of the acoustic impedance of the different material layers. The resolution of the measurement depends on the frequency of the incident wave; higher frequency waves provide for better resolution, but also result in more attenuation, effectively decreasing the penetration depth inside the material. In order to probe the material sufficiently deep with good resolution, one has to select an optimum frequency. The entire experimental setup is shown in Figure 2.\par

The sensitivity of the measurement depends on factors such as, attenuation of the wave (which is a function of frequency) and focal length of the transducer. The choice of the transducer is an important factor for getting maximum resolution. We chose the frequency of 15 MHz and a focal length of 2 inches for our measurements.\par

The data from the acoustic scan can be visualized in three different modes, 1) A-Scan, showing a digital waveform from measuring the reflected signal at a given location on sample 2) B- Scan, a collection of A-Scan amplitudes  along a chosen line resulting in a cross-sectional view of the sample,  3) C-Scan, a 2-D raster scan of A-Scans is performed to create  a 2-D image of the inside of the chips at a given depth. A specific and depth and depth range is achieved by integrating the A-scans over a corresponding time window for each point of the raster scan.\par

 The ASIC chips tested here are about 1 mm thick  and  a dimension of  2$\times$2 cm. The circuit made from 180 nm CMOS technology (1.8 V) includes about 320,000 transistors.\par 
 
The beam profile of the raster scan has a width of 250 $\mu m$ and the step size is 125 $\mu m$, which ensures that every point on the AISCs is scanned without missing a single point on the sample. In order to minimize errors on the measurement due to positional variations, a fixture was made to hold the ASICs in place.

\section*{THE CORRELATION ANALYSIS STRATEGY FOR FEATURE DETECTION}
\vskip.10 in
A correlation analysis has been developed to study the similarities in the chips’ internal materials structure within a group of ASICs. The signals are collected from each and every point on an ASIC, the scans yield a 150 x150 matrix of A-Scans, containing 22,500 pixels/chip in total.  In order to compare two ASIC chips for any structural differences, we calculate a cross-correlation value for each pixel based on the two corresponding A-scans by quantifying the similarity of their time traces over a given time window. Effectively we compare two ASIC chips at each position and thereby generate a 2-D image.  The main advantage of this analysis technique is that the correlation can be performed by choosing any particular time gate. A particular chosen time-gate allows us to perform a correlation analysis at any depth inside the ASICs. A low correlation value from the outside pins of ASICs are expected due to the distortions of the reflections from the edges (edge-effect) [8]. 

\begin{equation}
\Large{ \Phi_{fg}} = \frac{{}\sum_{i=1}^{n} f_{i}g_{i}}
{\sqrt{\sum_{i=1}^{n} f^{2}_{i}  \sum_{i=1}^{n} g^{2}_{i}}}
\end{equation}

The correlation coefficient for two different ASICs is defined according to equation (1). Here $f_{i}$, $g_{i}$ refer to the reflected waveforms in time bin i detected with the transducer at the location on the two different ASICs. These waveforms are in digitized format, recorded with a sampling time scale of 40 ns, slightly oversampling the 15 MHz pulsed acoustic signals.   Correlation value $\Phi_{fg}$ quantifies the similarity between the two waveforms over a chosen time window, corresponding to n time samples, and a specific depth regime inside the chips. In this analysis we make sure that the starting time of all the incident waveforms are synchronized, to ensure that there is no offset – as would be indicated by a time lag between the incident waves reflected off the top surface of the sample.  The resulting correlation values $\Phi_{fg}$ computed yield a 2-D map/image of the chip comparison. If the reflected waveforms of the two chips are close to identical, we expect a high correlation value close to 1. A correlation value of  0 indicates no correlation, due to substantially different waveforms. A correlation value of -1 corresponds to a phase shift between the waveforms from the two chips.  
\subsection*{EVALUATION OF UNCERTAINTY IN THE MEASUREMENT}
\begin{figure}[h]
\includegraphics[width=10cm]{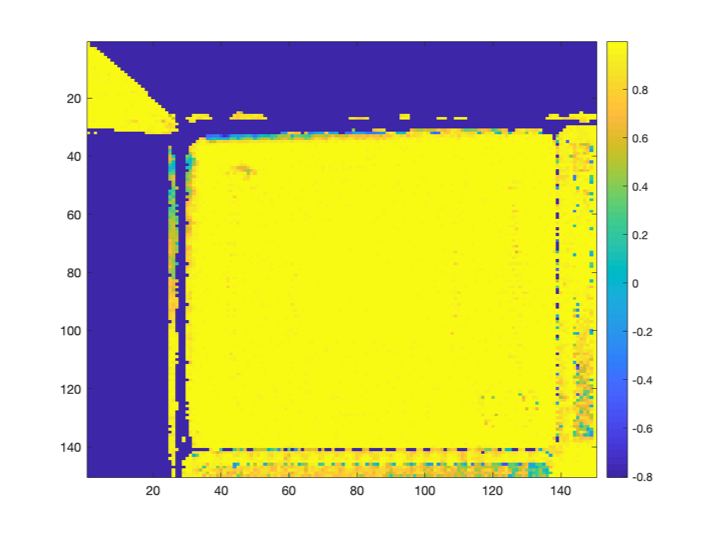}
\includegraphics[width=9.5cm]{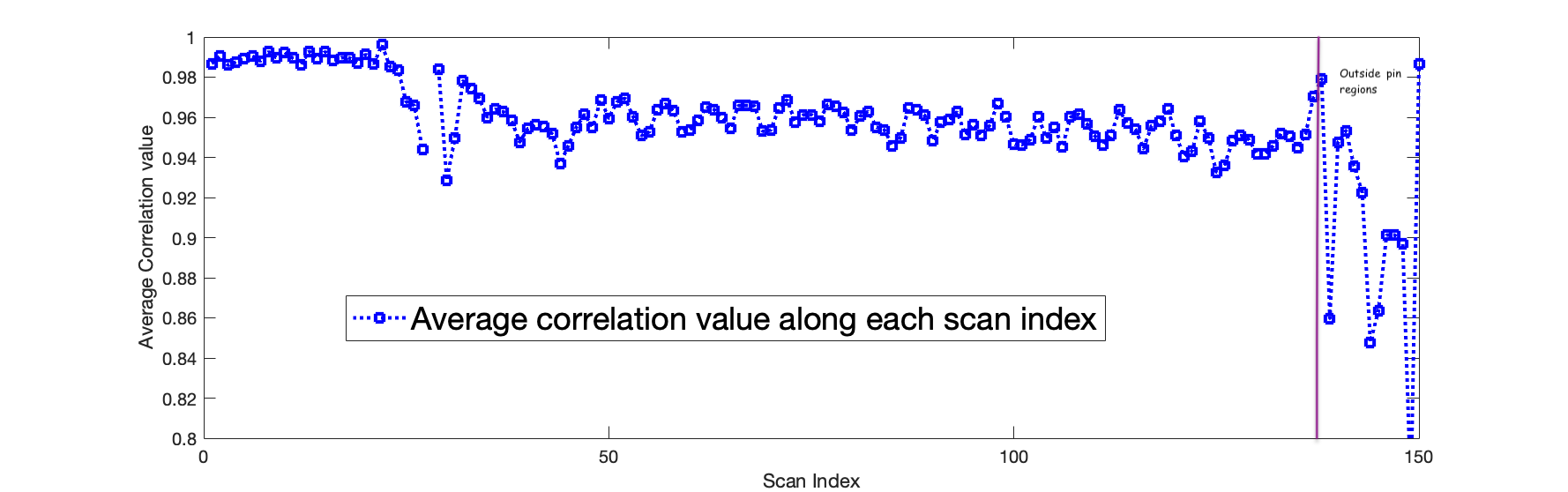}
\caption{ The top is the correlation of the same chip by taking after two consecutive measurement, and the bottom is the average value of  correlation along the scan indices. The Low correlation values corresponds  to the scattering from  the edges are due to the scattering of the acoustic wave.}
\label{sys}
\end{figure}

\vskip.10 in
The systematic uncertainty associated with the correlation measurement has been calculated by a measurement carried out on the same chip consecutively and by computing the correlation value, and the deviation from 1 is quoted as the error in the measurement. The uncertainty in the correlation measurement is about 5\%, see figure \ref{sys} (the outside pin region on the sample is not included due to the known effect of the scattering of the acoustic wave).

\section*{RESULTS AND DISCUSSION}
\vskip.10 in
The systematic uncertainty associated with the correlation measurement has been calculated by a measurement carried out on the same chip consecutively and by computing the correlation value, and the deviation from 1 is quoted as the error in the measurement. The uncertainty in the correlation measurement is about 5\%, see Figure \ref{sys} (the outside pin region on the sample is not included due to the known effect of the scattering of the acoustic wave). The similarity of the sample is evaluated by looking at the 2-D image formed using the correlation values. Images were created by forming the correlation values for the entire time window, and a select narrowed time-window to focus on a given depth inside the material. To get a sense of the sensitivity of this analysis, a set of ASICs was used to compare their inner structure to one reference chip.\par

Figure 5, shows the image formed using the correlation values for two different ASICs. To check for artifacts associated with a particular pixel size, we also performed the same study by averaging the correlation values over a pixel size three times the nominal pixilation (125 microns). As a result, a substantial number of  ASICs show a diagonal feature in the correlation value images. The time integration window chosen provides an estimate of the depth of this feature, and we conclude that this feature appears to coincide with the depth of the thermal paste-layer of those ASICs.  

\begin{figure}[h]
\includegraphics[width=4.25cm]{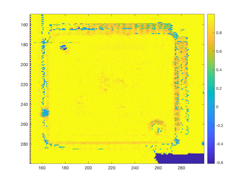}
\includegraphics[width=4.25cm]{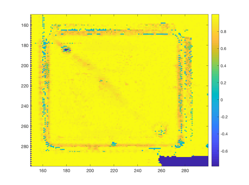}
\caption{ Images of the correlation analysis applied to 3 different ASICs are shown. To the left, we show the correlation map of two similar ASICs. It can be seen that there are no substantial differences inside the packaging (plastic packaging) of the chips. The smaller correlation values at the edges correspond to the pins and as it would be difficult to align those to the level of the imaging resolution. To the right, we show two ASICs that show substantial differences along a diagonal line, related to a difference inside the packaging of the circuitry.}
\end{figure}

The correlation analysis was also done in the frequency domain by taking the Fourier transform of the signal, and the features appeared to be the same.

A more general view of a collection of 9 ASICs and their internal structural similarities can be gleaned from Figure 6. Here, a distribution of the correlation values is shown for 8 chips compared to a reference chip.  As can be seen, most chips show correlation value distributions clustered around 1, corresponding to a high level of similarity of the inner structure when compared with the reference chip. One sample (chip identification number 1745) exhibits a very broad correlation value distribution.  This chip had been purposely exposed to mechanical pressure to induce delamination inside. Furthermore, ASIC 1732 shows a broadened correlation value distribution compared to the other chips, indicating that the inner structure also differs substantially from the reference chip.  

\begin{figure}[h]
\includegraphics[width=8cm]{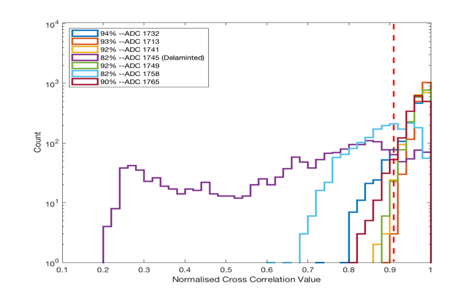}
\caption{The distribution of the correlation values for the damaged ASICs is compared with the other ASICs, a clear shift  towards the lower correlation value is noticed.}
\end{figure}

One of the key motivations for this study is to determine if any structural changes inside the ASICs’ composite material occur as a result of temperature cycling between room temperature and submersion in liquid nitrogen.  In these tests, several ASIC chips were immersed in liquid nitrogen followed by an acoustic microscopy scan done immediately after taken out. The summed amplitudes of the acoustic scans (C-scan) for one chip comparing the correlation images of two ASICs before (left) and after (right) the temperature cycling is shown in  Figure 7.  There is an apparent difference in the internal structure as indicated in Figure 7, to the left. A small strip feature appeared at a depth of 0.4 mm inside the ASIC package (calculated from the time of the reflected wave, the velocity of the sound in this composite material is taken as 2800 m/s) [9]. This mode of feautre inside the ASICs can be explained by the stitch crack caused by delamination [10], here in order to confirm we need to chemically open up the package and also to study the variations in the functional test. To gain a more detailed view of the regions of interest, we show in Figure 9 the corresponding B-Scan images of the chip before (top) and after the liquid nitrogen treatment (bottom). The comparison shows a clear change in the center region. It appears that a multi-layer indentation was already present before the temperature cycling (top), and is much enlarged after the cycling (bottom). To extract more information about the low correlation region, the pure waveforms from this region is compared before and after the liquid nitrogen treatment, we have observed a phase inversion in the waveform at a particular depth inside the chip, this is shown in Figure 8.
\begin{figure}[h]
\includegraphics[width=9cm]{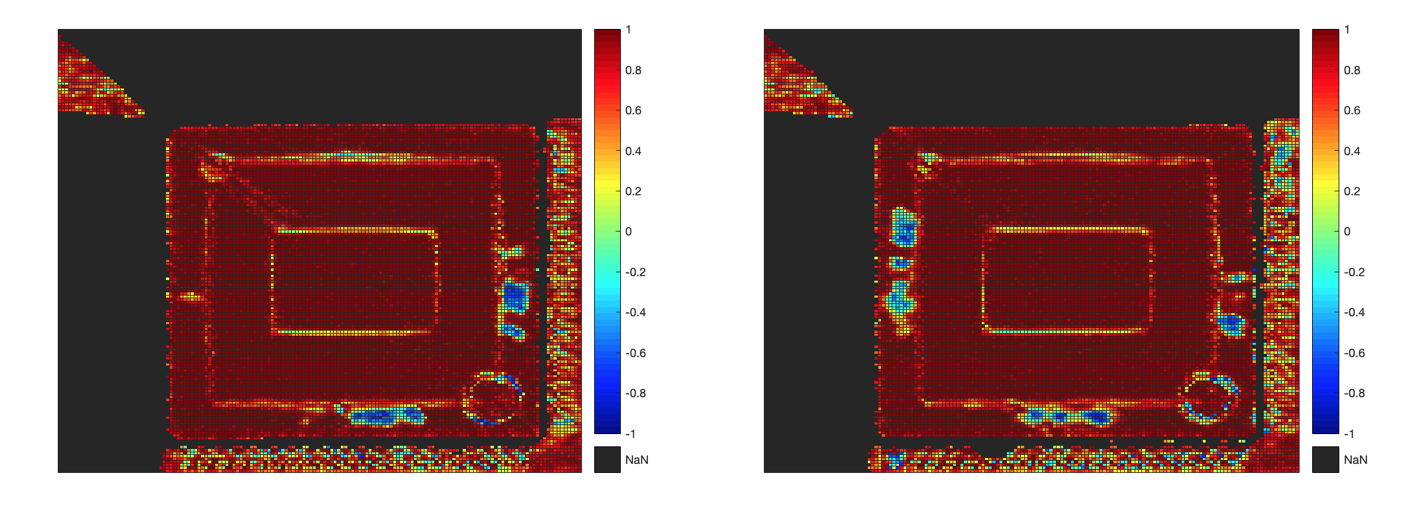}
\caption{The C-scan image  of two the ASICs formed using the correlation values of the samples before and after treated with liquid nitrogen, a clear change is observed in the sample. We have noticed these changes in a few of the samples. This image corresponds to a particular depth inside the sample. The waveform from the blue region shows a clear phase inversion. }
\end{figure}

\begin{figure}[h]
\includegraphics[width=9cm]{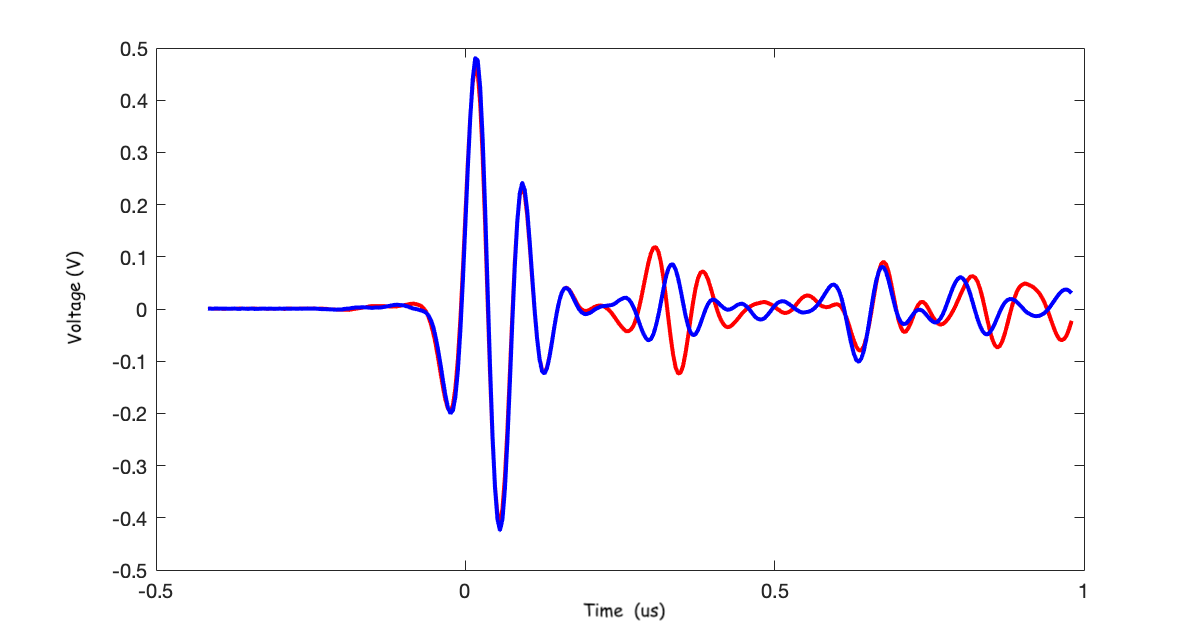}
\caption{The waveforms are compared after and before treating with the liquid nitrogen at the low correlation regions,  the clear phase inversion in the waveform, which is typically considered as an defect in the SAM (corresponding to the blue region in Figure 7).}
\end{figure}

\begin{figure}[h]
\includegraphics[width=9cm]{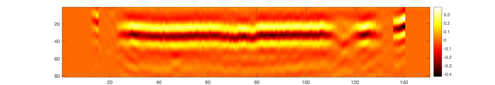}
\includegraphics[width=9cm]{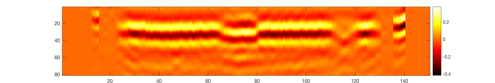}
\caption{The B-scan image from one of the ASICs before (top) and after(bottom) treated with the liquid nitrogen for 20 minutes. A shift in the B-scan waveform could be an indication of sensitive regions inside the chip package.}
\end{figure}
\section*{CONCLUSIONS}
\vskip.10 in
 Scanning Acoustic Microscopy is used here to evaluate its potential for ASIC chip non-destructive testing.  In this work, we have developed a correlation analysis that allows one to perform a comparison between different ASIC samples. We have demonstrated that this technique is highly sensitive to even small changes.
 For example, one can compare statistical samples of chips with different testing histories, such as failed functional tests and well-performing chips. In some of the chip samples, we were able to identify a diagonal feature across the inner chip at the depth of the thermal paste layer. 

Furthermore, this correlation analysis was used to study the effects of temperature cycling between 87K and room temperature.  Our results demonstrate structural changes in several chips that occurred after cryogenic cycling.  Additional studies are underway to evaluate the impact of these structural changes including X-ray imaging of wire bonds.  Finally, we are also planning to perform functional tests and destructive evaluation to fully understand the origin of these structural changes.

In the future, we will be developing a method, where we can do these measurements in the liquid nitrogen itself, which will a more realistic scenario as our detector taking data while the chip is immersed in the liquid nitrogen.  

\bibliographystyle{asmems4}

\begin{acknowledgment}
\vskip.10 in
This research is supported by grants from the U.S. Department of Energy Office of Science and by Iowa State University. We would like to acknowledge our colleagues Matt Worcester, Steve Kettel and Hucheng Chen from Brookhaven National Laboratory for providing ASIC chips samples and for the fruitful discussions of this work. 
\end{acknowledgment}

%
% ...

\bibliography{asme2e}


[1] B. Abi, et.al. (2017) “The single phase ProtoDUNE technical review report” arXiv:1706.07081 \\

[2] S. Gao, et.al. (2017), “The Development of Front-End Readout Electronics for ProtoDUNE-SP LAr TPC” Proceedings of Science Vol 313.\\

[3] B. Abi, et.al. (2017) ``DUNE TDR Deep Underground Neutrino Experiment (DUNE)" arXiv:1807.10334v1 \\

[4] T. Chen, et.al (2006) “CMOS reliability issues for emerging cryogenic lunar electronics applications,” Solid-State Electron., vol. 50, pp. 959–963 \\

[5] C. Hu, et.al “Hot-electron-induced MOSFET degradation-model, monitor, and improvement,” IEEE J. Solid-State Circuits, vol. SSC-20, no. 1, pp. 295–305, Feb. 1985. \\

[6] J. Yang. (1996) “Non-destructive identification of defects in integrated circuit package by scanning acoustic microscopy” Microelectron.Reliab.Vol.36, N0.9. pp.1291-1295,1996 \\

[7] Sandeep Kumar Diwendi, et.al. (2018) “Advances and Researches on Non-Destructive Testing: A Review” Materials Today: Proceedings 5 (2018) 3690–3698 \\

[8] Pouria Aryan et.al (2018) “An Overview of Non-Destructive Testing Methods for Integrated Circuit Packaging Inspection” Sensors 2018, 18, 1981  \\

[9] Toshio KONDO and Mituyoshi KITATUJI (2004) “Composite Materials and Measurement of Their Acoustic Properties” Japanese Journal of Applied Physics Vol. 43, No. 5B, 2004, pp. 2914–2915 \\

[10] M. van Soestbergen,A. Mavinkurve, S. Shantaram, J.J.M. Zaal (2017) ``Delamination-induced stitch crack of copper wires" IEEE 18 th Conference on Conference on Thermal, Mechanical and Multi-Physics Simulation and Experiments in Microelectronics and Microsystems (EuroSimE).


\end{document}